\documentclass[aps,prb,twocolumn,superscriptaddress,raggedbottom,showpacs]{revtex4}
\usepackage{graphicx,latexsym}
\usepackage{amsmath}
\usepackage{dcolumn}
\usepackage{bm}
\begin{document}
\title
{\bf Transient regime in non-linear transport through many-level quantum dots}
\author{Valeriu Moldoveanu}
\affiliation{National Institute of Materials Physics, P.O. Box MG-7,
Bucharest-Magurele, Romania}
\author{Vidar Gudmundsson}
\affiliation{Science Institute, University of Iceland, Dunhaga 3, IS-107 Reykjavik, Iceland}
\author{Andrei Manolescu}
\affiliation{National Institute of Materials Physics, P.O. Box MG-7,
Bucharest-Magurele, Romania}

\begin{abstract} We investigate the nonstationary electronic 
transport in noninteracting nanostructures driven by a finite bias and time-dependent signals 
applied
at their contacts to the leads. The systems are modelled by a tight-binding Hamiltonian
and the transient currents are computed from the non-equilibrium 
Green-Keldysh formalism. 
The numerical implementation is not restricted to weak coupling to the leads and 
does not imply the wide-band 
limit assumption for the spectral width of the leads. As an application of the method 
we study in detail the
transient behavior and the charge dynamics in single and double quantum dots
connected to leads by a step-like potential, but the method allows as well the consideration of
 non-periodic potentials or short pulses.  
We show that when the higher energy levels of the isolated system are located within the bias window
of the leads the transient current approaches the steady state in a non-oscillatory smooth fashion. 
At moderate coupling to the leads and fixed
bias the transient acquires a step-like structure, the length of the steps increasing 
with the system size. The number of levels inside a finite bias window can be tuned by a constant
gate potential.      
We find also that the transient behavior depends 
on the specific way of coupling the leads  to the mesoscopic system. 
\end{abstract}

\pacs{73.23.Hk, 85.35.Ds, 85.35.Be, 73.21.La}

\maketitle

\section{Introduction}

The dynamics of conduction electrons in open nanostructures modulated by time-dependent
signals is an outstanding problem
in quantum transport theory. Extensive experimental and theoretical work has been done
especially on quantum pumping \cite{Switkes,Geerligs,Brower,Levinson,Buttiker1,Arrachea}
(a detailed bibliography can be found in the recent review \cite{AM}) and photon-assisted tunneling
(see Ref. \onlinecite{Aguado}
and references therein). In these phenomena
one is interested in measuring or computing the current response of a mesoscopic system
driven by time-dependent {\it periodic} signals applied either on the system or on the attached leads.
In particular, an unbiased system subjected to two periodic potentials differing by a phase lag
generates a nonvanishing pumped current, provided one averages over the relevant period. 
\cite{Levinson}
If the signal frequency is small the pumping is adiabatic and can be described by a
'frozen' scattering matrix as is rigorously shown in Ref.\ \onlinecite{Avron1}. A photon-assisted tunneling
implies instead high frequencies and the measured current displays satellite peaks due 
to the additional 
side-bands. \cite{Kou}

From the theoretical point of view, any calculation of the current starts from defining the 
initial equilibrium state of the system and the perturbation that drives it. 
One way is to start with the connected system in the absence of the bias, and then to 
apply the bias adiabatically performing linear response calculations for the steady state current.
An alternative picture was proposed by Caroli \cite{Caroli} that takes as the equilibrium state the
state of the decoupled system with the bias already imposed on the leads. The perturbation here is instead 
the coupling to the leads 
that is usually adiabatically switched in the remote past and
eventually reaches its full magnitude at $t=0$. If one assumes that a steady state is
achieved, the Green functions depend only on time differences and the Keldsyh formalism 
gives the corresponding current in terms of Fourier transformed quantities. 
Both pictures were shown to be useful
in capturing and explaining important effects.   

As for an ac signal, it can be included in the Green-Keldysh approach as a time-dependent 
global shift of the 
spectrum of the leads. \cite{Jauho}
Note that the occupation probability (i.e. the Fermi function) of the leads stays
time-independent. Therefore this procedure assumes somehow that
an ac signal is applied as well adiabatically. In the quantum pumping calculations the
pumping potential is applied to the system in a steady-state, be it subjected
to a finite bias or not. Finally the current (transient, time-averaged or steady-state) is computed
from the Keldysh-Green function formalism. \cite{Haug}  

Here we aim to get some insight into a related topic: the calculation of the
transient current through a quantum dot (QD) whose coupling to the leads is time dependent while  
the bias applied on the leads is constant. 
There are considerably fewer theoretical results on
this issue (see the references below) and we were motivated also by recent
increasing interest in using suitable electric pulses to 
investigate relaxation processes in quantum dots by using pump and probe measurements or transient current
spectroscopy. \cite{Tarucha} As in the well known case of a turnstile pump \cite{TSP} these techniques 
imply oscillating tunnel barriers so that the transport formalism should deal with the nonlocal time-dependent 
coupling between the leads and the system.    

The problem we want to look at is defined as follows: i) The system is disconnected at any time
$t<0$, and the leads are submitted to a constant bias which is included through the
difference between the chemical potentials of the
leads. ii) At $t=0$ the leads are {\it suddenly} plugged to the system. Physically
 this means that the tunneling barriers between the leads and the system are set to be very
high at $t<0$ and drop suddenly at $t=0$ to an intermediate value that allows charge transfer
across the system. The simplest case of a constant barrier height at any $t>0$
opens already the problem of the existence of non-equilibrium steady-states in the long time limit.
In general and at a rigorous level, one can prove the existence of such states  when $t\to\infty$ \cite{Pillet} 
and moreover, a Landauer formula was shown to hold for the steady-state current. \cite{Nenciu}  
The method we developed is able to check the passage from transient behavior to steady state regime 
for specific systems, like many-level one- and two-dimensional quantum dots.
As we shall see, the onset of the steady state for a given system depends on the 
its structure (number of levels), on the measurement setup 
(the strength of the coupling to the leads and the location of the contacts), and also on
external parameters like gate potentials. 

Although, in the present work the numerical simulations are restricted to the step-like coupling to the leads, 
our model allows  
the consideration of more general time-dependent
potentials between the leads and the central region.
In particular we can investigate the response of a system to {\it nonlocal}
time-dependent perturbations that can be
switched on and off individually. 

Recently there have been several theoretical approaches to the transient
regime. In Ref. \onlinecite{Stefanucci} the time-dependent density
functional was used to compute the transient current in a one-dimensional 
system submitted to a finite bias applied on the leads. The coupling term does not depend on 
time and the system is in an equilibrium state in the absence of the bias. 
Starting from this state the Kohn-Sham equation is used to calculate the response of the system
to the external bias.
The same techniques allow the calculation of time-averaged
current in one-dimensional quantum pumps. \cite{Stefanucci1}  On the other hand,
Maciejko {\it et al.} \cite{Maciejko} have computed within the
Keldysh framework the response of a single-site dot for a
step-like or periodic signal applied to the {\it leads}, beyond
the wide band limit. In their approach the computation of the time-dependent Green 
functions of the perturbed system
uses steady-state Green functions of the coupled and biased system that have to be provided from 
density functional theory. We believe that theoretical results regarding the transient 
regime induced by time-dependent couplings of many-level systems would complement
these results.

The content of the paper is organized as follows: Section II 
presents the model and the theoretical tools we
use to compute the transient current. We rely essentially on the
non-equilibrium Green-Keldysh formalism. However, in contrast to
most of the previous studies we allow a complex structure for the
central region coupled to leads (i.e. there is more than one single
localized level and the system can be as complicated as we want: a
single dot, a double dot or an Aharanov-Bohm interferometer). Also
we go beyond the wide-band limit approximation and we solve exactly
the integral Dyson equation for the retarded Green function of the {\it coupled} 
central region by a
suitable numerical procedure. 
In doing so we take into account {\it all} the 
scattering processes between the leads and the sample. 

Although the electron-electron interaction could presumably 
play an important role in the transient behavior and the formalism we use allows the inclusion 
of Coulomb terms in the Hamiltonian, we do not take it into account in this work.
As is well known, the problem of the Coulomb interaction in the Keldysh approach is mainly technical and
implies suitable approximation schemes for the interaction self-energy. We postpone this issue for future work. Section III gives extensive discussion
of the numerical simulations for single and double dots. Section IV 
concludes the paper.

\section{Formalism}

The systems we study in this work have a typical transport configuration: 
a central region ($S$) coupled to two 
semiinfinite leads ($\alpha$ and $\beta$) via a tunneling term (see Fig.\,1 for a schematical representation). We shall use a tight-binding (TB) description of
the Hamiltonian which has the following form:
\begin{equation}
H(t)=H_S+H_L+H_T(t),
\end{equation}
where $H_S$ describes the system, $H_L$ the semiinfinite leads and $H_T(t)$ is the time-dependent
tunneling term:
\begin{equation}
H_T(t)=\sum_{\gamma=\alpha,\beta}\sum_{i\in\gamma}\sum_{m\in
S}V_{im}(t) (c_i^{\dagger}d_{m}+h.c).
\end{equation}
\begin{figure}[tbhp!]
\includegraphics[width=0.45\textwidth]{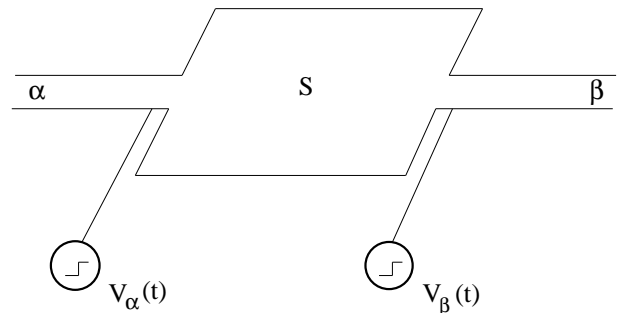}
\caption{Schematic picture of the system. The step-like potential is applied
between the leads and the central region $S$ at $t=0$.}
\label{figure1}
\end{figure}
Here $c_i$ and $c_i^{\dagger}$ denote the annihilation/creation
operators on the $i$-th site of the lead $\gamma$.
Similarly $d_m$ and $d_m^{\dagger}$ is the pair of
operators corresponding to the $m$-th site from the central region $S$. $
V_{im}(t)$ is the time-dependent hopping coefficient between the $i$-th site of the 
lead $\gamma$ 
and the $m$-th site of the central region. We take
here a nearest-neighbor coupling so the double sums in the above
expression contain only pairs of sites from the leads endpoint and the corresponding
contact region from the central system:
\begin{eqnarray}\label{cor1}
V_{im}(t)=\left\lbrace \begin{array}{ccc}
V_{\gamma}(t)\qquad {\rm if}\, i,m \quad {\rm nearest\, neighbors}\, \\
0 \qquad {\rm otherwise }
\end{array}\right.
\end{eqnarray}
In this work we consider a steplike potential, i.e. $V_{\gamma}(t)=V_{\gamma}$ if $t>0$ and 
zero otherwise.
$H_S$ has a usual tight-binding form
\begin{equation}
H_S=\sum_{m=1}^{N}(\epsilon_m+V_g)d_m^{\dagger}d_m+\sum_{\langle m,n\rangle }t_{mn}d_m^{\dagger}d_n.
\end{equation}

Here $t_{mn}$ are hopping terms, 
$\langle m,n\rangle $ denotes nearest-neighbor summation over the
system sites. $\epsilon_m$ is the on-site energy and the diagonal
term $V_g$ simulates a plunger gate potential applied on the system.
$N$ is the number of sites in the central region.
 The spectral width of the tight-binding lead is as usual $w:=[-2t_L,2t_L]$, where $t_L$ is 
the hopping energy on leads (we take the same hopping constant on every lead). 
In the numerical calculation we choose $t_L$ such that it covers entirely the
 spectrum of the central region but we do {\it not} assume it to be infinite, as it is done in the
wide-band limit approximation.

The central problem in electronic transport is to compute the statistical average of the
time-dependent current operator in a given lead (say $\alpha$)
$J_{\alpha}(t)={\rm Tr}\{\rho(t)j_{\alpha}(t)\}$ using the statistical operator $\rho(t)$. 
Notice that the time-dependence of the current
operator appears only because of the time-dependent coupling. Denoting by
$\{i_{\alpha}\}$ the endpoint sites of the lead $\alpha$ which are coupled to
the sites $\{m_{\alpha}\}$ of the central region and by $M$ the number
of sites in the transverse direction of the lead, the current operator
$j_{\alpha}$ has the form (we take the electron charge as $-e$ and $e>0$):
\begin{equation}
j_{\alpha}(t)=\frac{ie}{\hbar}\sum_{i_{\alpha}=1}^M
\sum_{m_{\alpha}\in C_{\alpha}}
V_{i_{\alpha}m_{\alpha}}(t)(c_{i_{\alpha}}^{\dagger}d_{m_{\alpha}}-d_{m_{\alpha}}^{\dagger}c_{i_{\alpha}}).
\end{equation}
Since the statistical operator $\rho(t)$ of the coupled system is not
easy to compute it is useful to move the time-dependence entirely to 
the current operator by writing $\rho(t)$ in terms of the equilibrium statistical operator $\rho_0$ 
of the disconnected system. In general, the coupling to the leads is established at a given instant
$t_0$ so that $\rho(t)=\rho_0$ for $t<t_0$. Then using the unitary evolution $\tilde U$ 
of the full Hamiltonian in 
the interaction picture w.r.t.\, the unperturbed one the solution of the quantum Liouville equation is given
by
\begin{equation}
\rho(t)=e^{-it(H_S+H_L)}{\tilde U}(t,t_0)\rho(t_0){\tilde U}(t,t_0)^*e^{it(H_S+H_L)}
\end{equation} 
Then it can be shown (see e.g. Ref. \onlinecite{Rammer}) that:
\begin{equation}
J_{\alpha}(t)={\rm Tr}\{\rho_0T_C(e^{-i\int_Cds{\tilde
H}_T(s)}{\tilde j}_{\alpha}(t))\},
\end{equation}
where $T_C$ is the ordering operator on the Schwinger-Keldysh contour
$C$ that runs from $t_0$ to $t$ and back to $t_0$. We remind the reader that in the
case of adiabatic coupling the statistical operator becomes
time-independent only in the remote past $t_0\to -\infty$. Both the
coupling and current operators are written in the interaction
picture.
%\begin{eqnarray}\nonumber
%{\tilde j}_{\alpha}(t)&=&e^{it(H_S+H_L)}j_{\alpha}(t)e^{-it(H_S+H_L)}\\
%{\tilde H}_T(t)&=&e^{it(H_S+H_L)}H_T(t)e^{-it(H_S+H_L)}.
%\end{eqnarray}
Using the definitions of the lesser Green functions in terms of the
Heisenberg operators:
\begin{eqnarray}\nonumber
G^<_{m_{\alpha}i_{\alpha}}(t,t')&=&i\langle c^{\dagger}_{i_{\alpha}}(t')d_{m_{\alpha}}(t)\rangle\\
G^<_{i_{\alpha}m_{\alpha}}(t,t')&=&i\langle d^{\dagger}_{m_{\alpha}}(t')c_{i_{\alpha}}(t)\rangle,
\end{eqnarray}
it follows that the current is given by a simpler relation:
\begin{equation}
J_{\alpha}(t)=\frac{2e}{\hbar}\sum_{i_{\alpha}=1}^M\sum_{m\in
C_{\alpha}}{\rm Re}(V_{i_{\alpha}m_{\alpha}}(t)G^<_{m_{\alpha}i_{\alpha}}(t,t)).
\end{equation}
 At this point the standard Keldysh formalism requires the application of the so called
 Langreth rules \cite{Haug} in order to express the lesser Green function
 $G^<_{m_{\alpha}i_{\alpha}}$ in terms of the Green functions of the
 central region in the presence of the leads
 $G^{<,R}_{m_{\alpha}n_{\alpha}}$ and the Green functions of the
 isolated semiinfinite lead $g^{<,A}_{i_{\alpha}j_{\alpha}}$. The
 latter can be analytically computed:
\begin{widetext}
\begin{eqnarray}
g^A_{i_{\alpha},j_{\alpha}}(t,t')&=&i\theta(t'-t)\sum_{p=1}^M
\chi_p(i_{\alpha})\chi_p(j_{\alpha})\int_{-2t_L+E_p}^{2t_L+E_p}dE
\rho(E-E_p)e^{-iE(t-t')}\\
g^<_{i_{\alpha},j_{\alpha}}(t,t')&=&i\sum_{p=1}^M
\chi_p(i_{\alpha})\chi_p(j_{\alpha})\int_{-2t_L+E_p}^{2t_L+E_p}dE
\rho(E-E_p)e^{-iE(t-t')}f_{\alpha}(E).
\end{eqnarray}
\end{widetext}
In the above equations $E_p=2t_L\cos(p\pi/(M+1))$ is the energy of
the transverse channel $p$. These channels appear due to the width of the leads
which in the tight-binding description is given by the number of the sites $M$ in the transverse direction.
More exactly, the many channel lead is constructed by taking $M$ semiinfinite 1D leads and by coupling 
them through 
nearest neighbor hopping constants. Then 
$\chi_p(i_{\alpha})=\sqrt{\frac{2}{M+1}}\sin(\frac{pi_{\alpha}\pi}{M+1})$
is the transversal eigenfunction associated to $E_p$, and $\rho(E)$
is the density of states at the endpoint of a semiinfinite one-dimensional lead:
\begin{equation}
\rho(E)=\theta(2t_L-|E|) \frac{\sqrt{4t_L^2-E^2}}{2t_L^2}.
\end{equation}
Finally, $f_{\alpha}(E)$ is the Fermi function in the lead $\alpha$.
The bias is included in our approach as the difference between the
two chemical potentials of the leads $V=\mu_L-\mu_R$. 

Plugging all these elements in the current formula one gets the main
expression that will be numerically implemented in the next section:
\begin{widetext}
\begin{equation}\label{current}
J_{\alpha}(t)=-\frac{2e}{h}{\rm Im} (\sum_{p=1}^M
\sum_{m_{\alpha},n_{\alpha}\in
C_{\alpha}}\int_{-2t_L+E_p}^{2t_L+E_p}dE \int_{0}^tdse^{-iE(s-t)}
\Gamma^{\alpha,p}_{m_{\alpha},n_{\alpha}}(E;t,s)
(G_{m_{\alpha}n_{\alpha}}^R(t,s)
f_{\alpha}(E)+G_{m_{\alpha}n_{\alpha}}^<(t,s)) ). 
\end{equation}
\end{widetext}
We have introduced the energy and time-dependent quantity
$\Gamma^{\alpha,p}_{m,n}$ that takes also into account the $M$
channels in the lead:
%\begin{widetext}
\begin{eqnarray}\nonumber
\Gamma^{\alpha,p}_{m_{\alpha},n_{\alpha}}(E;t,s)=\sum_{i_{\alpha},j_{\alpha}=1}^M&&\rho(E-E_p)
\chi_p(i_{\alpha})\chi_p(j_{\alpha})\times \\ \label{gamma}
&&V_{i_{\alpha}m_{\alpha}}(t)V_{j_{\alpha}n_{\alpha}}(s).
\end{eqnarray}
%\end{widetext}
A similar formula can be written down for the current $J_{\beta}$
and therefore one defines as well the net current
\begin{equation}
J(t)=J_{\alpha}(t)+J_{\beta}(t).
\end{equation}
 It is important to observe that in contrast
to the simple case of a single-site system the expressions for the
two currents imply the Green functions at different contacts.
%In the
%case of {\it equal} couplings to the leads the two time-dependent
%linewidths $\Gamma_{\alpha,p}=\Gamma_{\beta,p}:=\Gamma_p$ and the
%net current reads as
%\begin{eqnarray}\nonumber
%J(t)=-\frac{2e}{h}{\rm Im} (\sum_{p=1}^M \int_{-2t_L+E_p}^{2t_L+E_p}dE
%\int_{0}^tdt_1 \Gamma_{p}(t_1,t,E)\times \\\label{current}
%\times (G_{m_{\alpha}m_{\alpha}}^R(t,t_1)
%f_{\alpha}(E)+G_{m_{\beta},m_{\beta}}^R(t,t_1)f_{\beta}(E)+
% G_{m_{\alpha}m_{\alpha}}^<(t,t_1))+G_{m_{\beta}m_{\beta}}^<(t,t_1)) ).
%\end{eqnarray}
 The retarded and the lesser Green functions are then to be computed from
 the Dyson and Keldysh equations \cite{Haug}:
\begin{widetext}
\begin{eqnarray}\label{Dyson}
G^R(t,t')&=&G^R_0(t,t')+\int_0^tdt_1G^R(t,t_1)\int_0^{t_1}dt_2
\Sigma^R(t_1,t_2)G^R_0(t_2,t')\\\label{Keldysh}
G^<(t,t')&=&\int_0^tdt_1G^R(t,t_1)\int_0^{t'}dt_2\Sigma^<(t_1,t_2)G^A(t_2,t'),
\end{eqnarray}
\end{widetext}
where $G^{R,A}_0(t,t')$ are the retarded and advanced Green
functions of the isolated central region and $\Sigma^{R,<}$ are the
retarded and lesser self-energies. $G^R_0(t,t')$ has a simple
expression in terms of the discrete spectrum $\{E_{\lambda}\}$ of
the central region and its localized eigenfunctions $\psi_{\lambda}$
(clearly $\lambda=1,..N$):
\begin{equation}
G^R_{0,mn}(t,t')=-i\theta(t-t')\sum_{\lambda}\psi_{\lambda}(m)
\overline{\psi_{\lambda}(n)}e^{iE_{\lambda}(t-t')}.
\end{equation}
We emphasize the lower integration limit $t=0$ in equations
(\ref{Dyson}) and (\ref{Keldysh}). This is due to the fact that
there is no coupling for $t<0$. In the adiabatic setup the coupling
is established in the remote past and one should set a lower cutoff
in the numerical implementation.
However, the Dyson equation still contains two coupled integrals.
The two self-energies above contain the information from the leads
and are finite rank matrices in the Hilbert space of the central
region $S$ ($\gamma=\alpha,\beta$):
\begin{eqnarray}\label{SigmaR}
\Sigma^R_{mn}(t,t')&=&\sum_{\gamma}V_{\gamma}(t)
g^R_{i_{\gamma},j_{\gamma}}(t,t')V_{\gamma}(t')\delta_{mm_{\gamma}}
\delta_{nn_{\gamma}}\\\label{SigmaM}
\Sigma^<_{mn}(t,t')&=&\sum_{\gamma}V_{\gamma}(t)
g^<_{i_{\gamma},j_{\gamma}}(t,t')V_{\gamma}(t')\delta_{mm_{\gamma}}\delta_{nn_{\gamma}}.
\end{eqnarray}
We stress that the indices of the leads' Green function are
unambiguously determined as the neighbor sites of the contact surface
$C_{\alpha}$.  In the single channel case $M=1$ one recovers simpler
expressions. In particular the retarded Green function of the lead
can be expressed through the Bessel function of the first kind:
\begin{equation}\label{gmic}
g^R_{1_\gamma,1_\gamma}(t,t')=\frac{-i\theta(t-t')J_1(2t_L(t-t'))}{2t_L(t-t')}.
\end{equation}
We point out the difference between the exact form of the retarded self-energy and
the simple wide-band limit expression (which simplifies to $\delta(t-t')$ up to some constants).
Note that the retarded Green function gives the leads' self-energy and is a highly oscillating functions.
it will turn out in Section III that this behavior has crucial effects on the transient current.
Another difficulty of Eq.(\ref{Dyson}) comes from the quadratic dependence
of the self-energies on the time-dependent coupling. Clearly this prevents any partial
Fourier transform trick.

Given these,  our strategy in solving the integral Dyson equation relies in transforming
it into an algebraic equation of the form $AX=B$ where $A,X,B$ are generalized complex matrices
depending on both spatial and time arguments. To this end we first plug the retarded self energy from
Eq.(\ref{SigmaR}) in the Dyson equation (\ref{Dyson}) and discretize the time arguments. Note that
the variable $t_2$ is defined on a denser grid than the one used for $t_1$.
The inner time integral is evaluated by a repeated 4-point Gauss method, which turned out to be accurate enough
for the numerical results to be stable when increasing the number of integration steps.
This procedure allows us to write the double integral as
a matrix $\tilde{G^R }\tilde{A}$, where $\tilde{A}$ is actually a product of
$G^R_0,\Sigma^R$ and some diagonal matrices containing the Gauss weights needed in the
integration procedure. Then the adjoint of the generalized retarded Green function
$\tilde{G^R }$ is simply the solution of the algebraic equation
$(1-\tilde{A})^*\tilde{G^R }^*=\tilde{G^R_0}^*$. The true Green function is recovered by
turning back the mixed indices of $\tilde{G^R }$. We stress that by solving the equation
for $\tilde{G^R }^*$ the Dyson equation is solved exactly. Moreover, no matrix inversion is required.
This is certainly an advantage in the numerical simulations since it is known that
matrix inversion is both memory and time-consuming. The advanced Green function is computed
using the identity $G^A_{ij}(t,t')=G^R_{ji}(t',t)$ and the lesser Green function is derived
from the Keldysh equation. Also, the time-dependent occupation number can be computed as:
\begin{equation}
N(t)={\rm Im}\sum_{m\in S}G^<_{mm}(t,t).
\end{equation}
The current in the right lead $J_{\beta}$ has a similar expression.
We note that for a system with many sites one has to deal with
different contact Green functions besides replacing only the Fermi
function in the first term in the current formula. Moreover, in the
transient regime the current conservation does not imply the usual
identity $J_{\alpha}=-J_{\beta}$. We shall discuss this feature
below.

In the Keldysh approach to time-dependent transport the problem is
to extract physical information from the two contributions in 
Eq.(\ref{current}). In the simplest case of a single-site and within
the wide-band limit it was shown that the {\it average} current
obeys a Landauer-like formula. The effects of a step-like or
harmonic time-dependent potentials applied adiabatically on leads
were studied both in the WBL \cite{Jauho} and beyond
\cite{Maciejko}. However, to our best knowledge no transient
current calculation for a many-level structure beyond the wide-band
limit has been performed within the Keldysh formalism.

\section{Numerical simulations}

In all the plots the bias, the energy, the hopping constants on the leads, the coupling strengths and 
the gate potentials will be expressed in 
terms of the hopping energy of the central region $t_D$ which is chosen as energy unit. 
The current is therefore 
given in units of $et_D/\hbar$ and the time expressed in units of $1/t_D$.
Since the spectrum of the two-dimensional discrete Laplacian covers the range $[-4t_D,4t_D]$ we shall 
take $t_L=2$ in order to match it to the spectral width of the one-dimensional lead $[-2t_L,2t_L]$.
We take also $e=\hbar=1$.
 The current given by 
Eq.\,(\ref{current}) can be written as a sum of two contributions
\begin{equation}
J_{\alpha}(t)=J_{\alpha}^R(t)+J_{\alpha}^<(t),
\end{equation} 
the '$<$' and '$R$' labelling emphasizing that the corresponding term 
contains the lesser and retarded Green functions. 
 We shall consider for simplicity only single channel leads.

\subsection{Single-site}

\begin{figure}[tbhp!]
\includegraphics[width=0.45\textwidth]{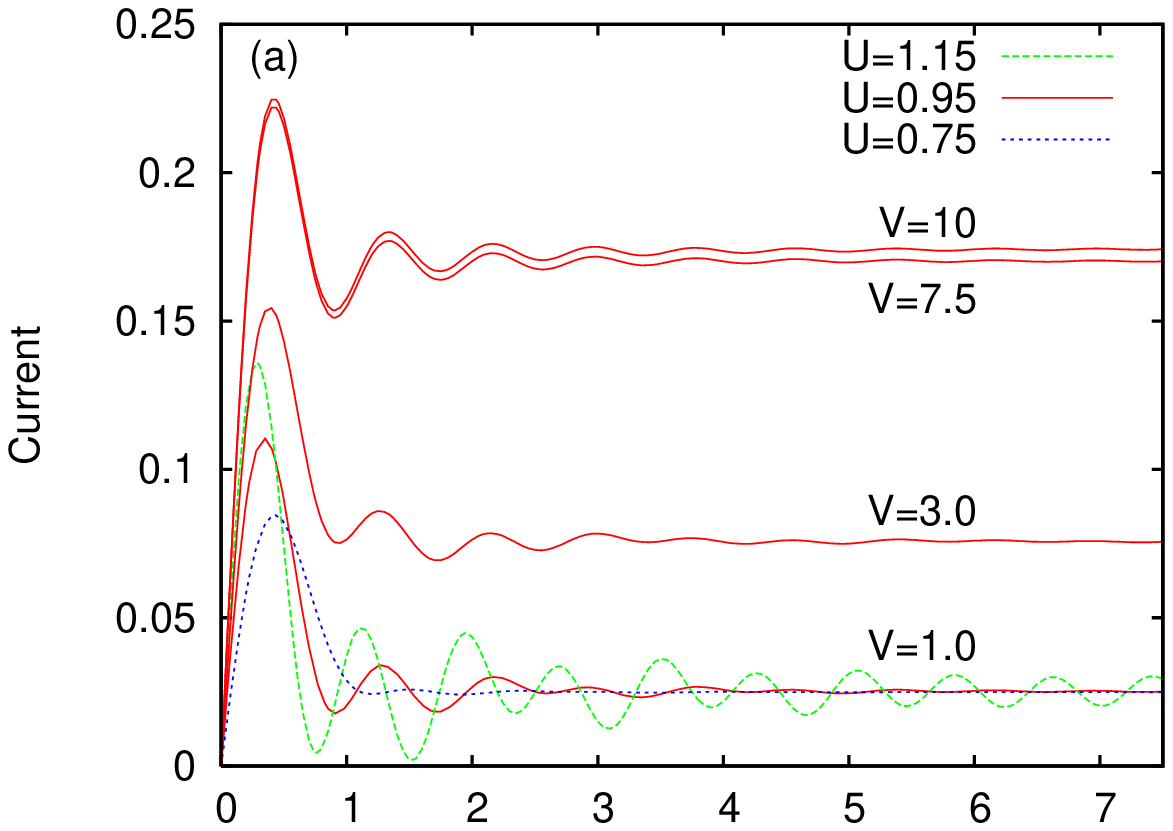}
\includegraphics[width=0.45\textwidth]{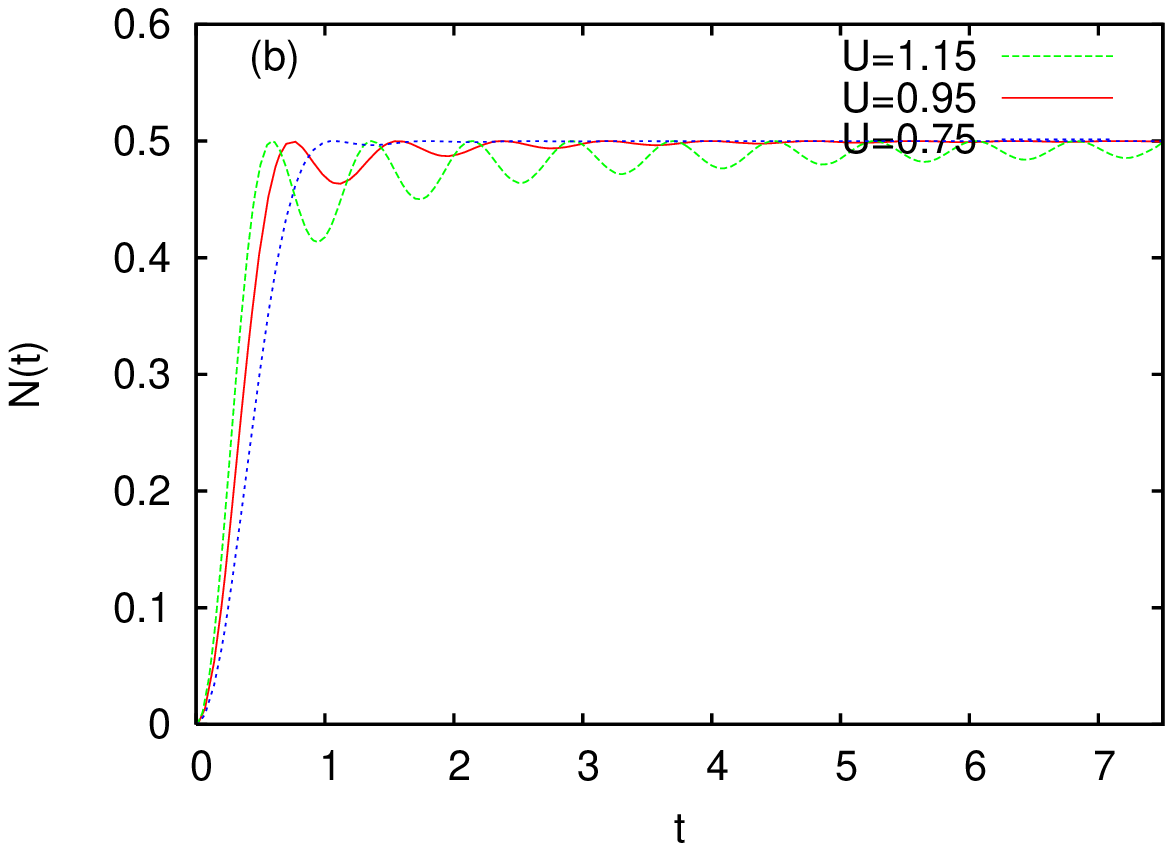}
\caption{(Color online) (a) The transient current for one site coupled to single-channel leads. We present curves
for different values of the
coupling strength $U$ and of the bias $V$. (b) The effect of the coupling amplitude $U$ on the
occupation number. The bias is fixed to $V=1.0$, and $kT=0.0001$.}
\label{figure2}
\end{figure}

We start this section by discussing the case of a single
site dot which allows qualitative discussion on the transient regime and on the transition towards the steady-state.
The site is coupled to single-channel leads (i.e. $N=M=1$). 
Both leads
are coupled suddenly to the system with the same strength $U$, i.e
$V_{\alpha}=V_{\beta}:=U$. The bias is applied symmetrically on leads, 
i.e. $\mu_{\alpha,\beta}=\mu_0\pm eV/2$,
$\mu_0$ being the chemical potential of the unbiased leads. We observe that for $\mu_0=0.0$ the single eigenvalue
of the isolated system $E_0=0.0$ is located in the middle of the bias window $W=[\mu_0-eV/2,\mu_0+eV/2]$. 
As we shall see later on, the position of the eigenvalues of the system within the bias window has important 
implications on the transient current.
The free retarded
Green function of the single site system is simply $G^R_0=-i$ 
but the full retarded Green function is still given by the {\it
integral} Dyson equation and an analytic solution is not at hand. 

Fig.\,2(a) gives the transient current for different values of the bias
$V$ and of the coupling amplitude $U$ and reveals that the
parameter that controls the shape and the amplitude of the
oscillation is the coupling strength $U$. At moderate coupling $U=0.75$ the
steady state (SS) is achieved fast but an oscillatory behavior is observed at
$U=0.95$. The case $U=1.15$ is 
beyond the perturbative regime and the steady-state is not achieved
in the selected time-range. 
As the bias increases 
the current saturates
for values of $V$ that exceed the spectrum of the leads (i.e. for
$V>8$), emphasizing the non linear transport regime. In
turn, the bias neither affects the amplitude nor the period
of the oscillations. This is due to the fact that in our model, as 
in all approaches based on the Keldysh formalism, there is
no term in the Hamiltonian to describe the voltage drop across the sample, the bias being included only via
the Fermi functions of the leads. 

\begin{figure}[tbhp!]
\includegraphics[width=0.45\textwidth]{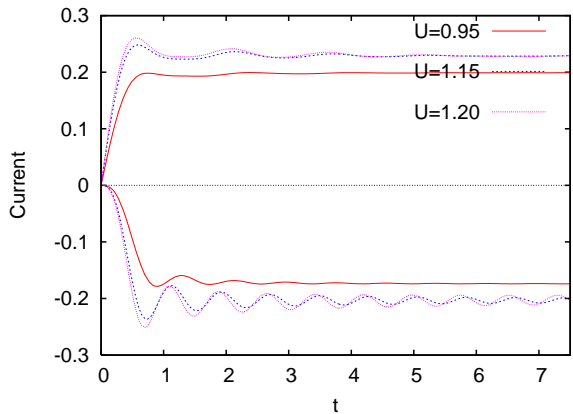}
\caption{(Color online) The two contributions to the transient current in the left lead. $J_{\alpha}^R(t)$
is always positive
while the lesser contribution $J_{\alpha}^<(t)$ is negative. The chosen values for the coupling strength $U$
are given in the figure. The bias $V=1.0$, and $kT=0.0001$.}
\label{figure3}
\end{figure}

In Fig.\,2(b) we plot the occupation
number of the resonant site $N(t)$ for the same parameters as in
Fig.\,1(a). The behavior w.r.t. $U$ is similar. Few more things worth
to be noticed: i) In the steady state regime the occupation of the site is 1/2; 
ii) as $U$ increases while the bias stays constant the
occupation number reaches its maximum value faster and the value
corresponding to the steady state decreases; 
 iii) Comparing
Fig.\,2(a) and 2(b) it can be seen that there is no clear relation
between the principal maximum of the current and the one of the occupation
number; in fact the electrons are accumulating in the system even
after the current starts to decrease towards the steady state.

Fig.\,3 shows the two contributions to the current $J_{\alpha}^R$
and $J_{\alpha}^<$ for $V=1.0$ and several coupling
constants considered in Fig.\,2. 
A physical significance of these two currents was proposed in
Ref. \onlinecite{Jauho} for the single site case. Although both Green functions at the contacts appearing
in the current formula
are 'dressed' by the leads' self-energy one could view $J_{\alpha}^R$ as the current flowing towards the sample and
$J_{\alpha}^<(t)$ as the current from to sample to the lead $\alpha$.
 One notices that the currents
have opposite signs. 
Another observation is that the lesser contribution is responsible
for the total current oscillations since $J_{\alpha}^R$ saturates
quickly. However at small times $J_{\alpha}^R$ grows faster than
$J_{\alpha}^<$, leading thus to the fast increase of the transient.

%In order to understand these features we have to analyze only the contact
%Green functions appearing in the current formula, because the oscillatory integral over energy
%does not depend on the coupling strength. 

In order to understand the nature of the oscillations in the transient current and their dependence on the 
coupling strength $U$ it is useful to rewrite the current formula
Eq.\,(\ref{current}) in a more useful form (since we
consider a single site system there is only one contact site and the
indices of the Green functions can be omitted):
%\begin{widetext}
\begin{equation}\label{partial}
J_{\alpha}(t)=-2U^2{\rm
Im}\int_0^tds(G^R(t,s)F_1(s,t)+G^<(t,s)F_2(s,t)),
\end{equation}
%\end{widetext}
where $F_1,F_2$ are two oscillating integrals:
\begin{eqnarray}\label{F1}
F_1(s,t)&=&\int_{-2t_L}^{2t_L}dEf_{\alpha}(E)\rho(E)e^{-iE(s-t)}\\\label{F2}
F_2(s,t)&=&\int_{-2t_L}^{2t_L}dE\rho(E)e^{-iE(s-t)}.
\end{eqnarray}
One can easily observe that actually $F_2(s,t)$ can be expressed
through Bessel function of the first kind:
\begin{equation}
F_2(s,t)=\frac{\theta(t-s)J_1(2t_L(t-s))}{2t_L(t-s)}.
\end{equation}
For fixed $t$, $F_2$ is an oscillating function of $s$ whose
oscillation amplitudes increase with $s$. $F_1$ does not have a
simple analitical expression but it has a similar behavior.

The oscillatory behavior of the current is clearly decided by the convolution in Eq.\,(\ref{partial}).  
Besides the oscillations of $F_1$ ans $F_2$ one expects as well a complex behavior of the Green functions.
We recall that the Dyson equation counts
the infinite back-and-forth tunneling processes involving the leads
and that the amplitudes of these events are even powers of $U$. Now, the higher order terms in
the Dyson equation contain multiple integrals of products of the leads' self-energy which is highly
oscillating (see (\ref{gmic}))
Therefore if $U<<1$ there will be only few low-order significant
contributions from the complicated lead-sample scattering. The critical value $U=1.00$
corresponds to the onset of the nonperturbative regime, and the method we use for solving the Dyson equation
captures as well this situation, taken into account all contributions. 

\begin{figure}[tbhp!]
\includegraphics[width=0.45\textwidth]{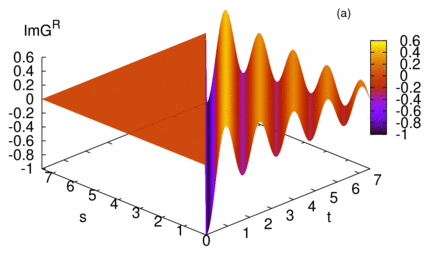}\\
\includegraphics[width=0.45\textwidth]{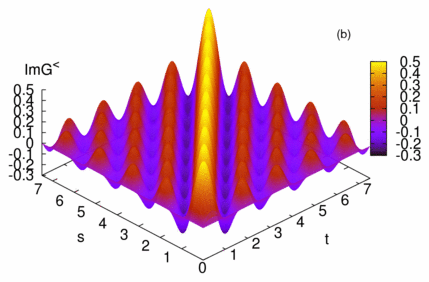}
\caption{(Color online) The imaginary parts of the retarded (a) and lesser (b) two-time Green functions of the
coupled single site. The bias
$V=1.0$ and the coupling strength $U=1.2$. The oscillations seen along the 'diagonal' in (b)
corresponding to almost equal times are responsible for the oscillations of the occupation number.}
\label{figure4}
\end{figure}

We give in Fig.\,4 the 3D plots of the imaginary parts
for the retarded and lesser Green functions at coupling strength $U=1.20$, which leads to oscillations
of the transient. These are the relevant quantities in the current formula 
since it turns out that
the real part of $G^R$ and of $F_2$ are vanishingly small (not shown). 

\begin{figure}[tbhp!]
\includegraphics[width=0.45\textwidth]{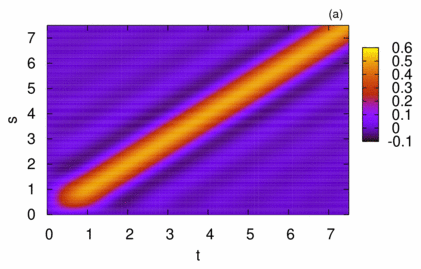}\\
\includegraphics[width=0.45\textwidth]{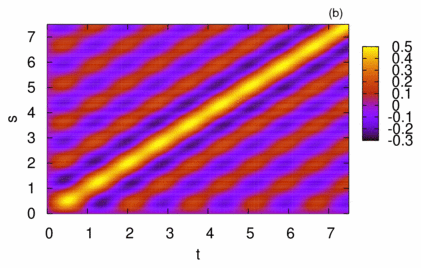}
\includegraphics[width=0.45\textwidth]{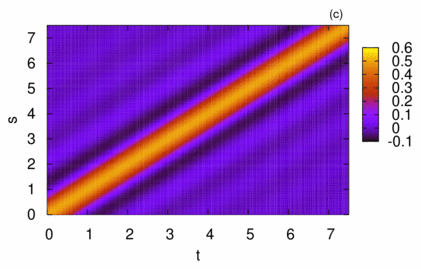}
\caption{(Color online) The imaginary part of the lesser two-time Green function
for the moderate coupling $U=0.75$ (a) and strong coupling $U=1.2$ (b). 
(c) The real part of the oscillating function $F_2$. The bias
$V=1.0$.} 
\label{figure5}
\end{figure}

\begin{figure}[tbhp!]
\includegraphics[width=0.45\textwidth]{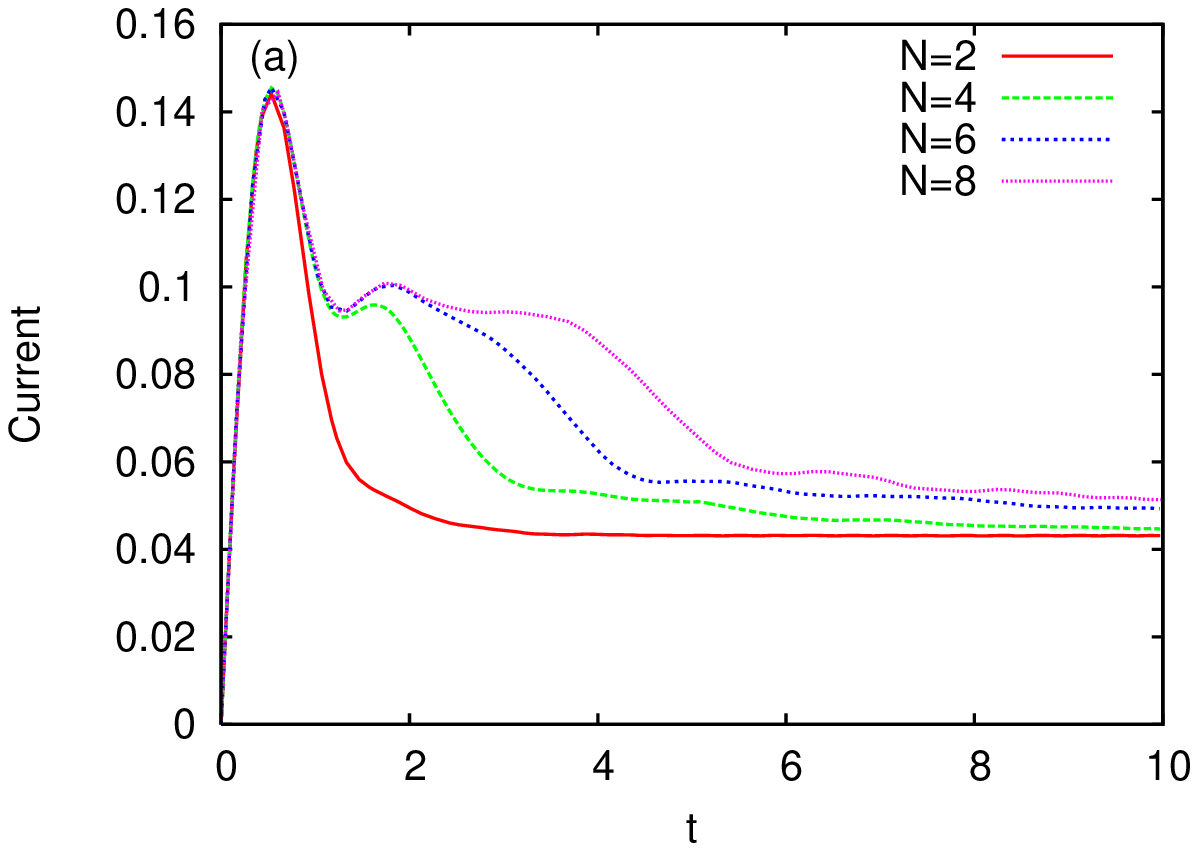}
\includegraphics[width=0.45\textwidth]{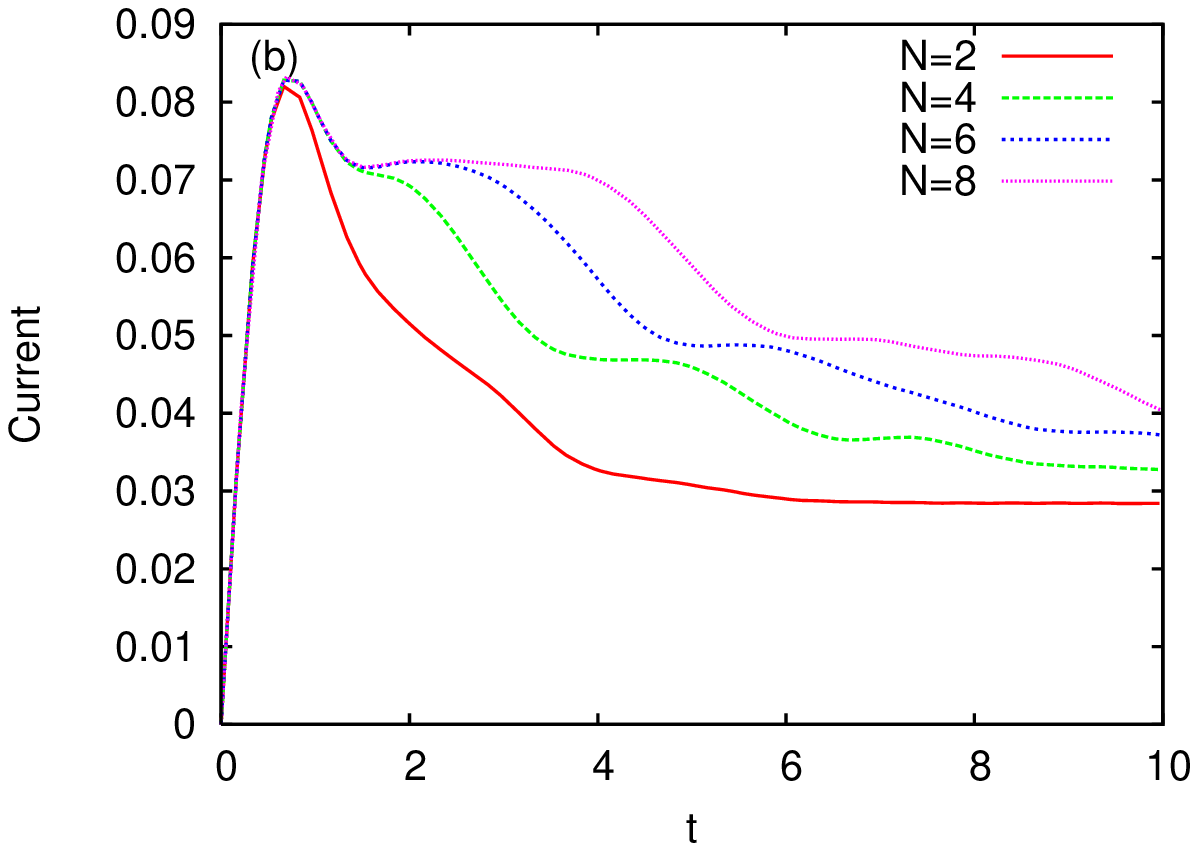}
\caption{(Color online) (a) The transient current in the left lead $J_{\alpha}(t)$
 for different sizes of the 1D central region.
The number of sites $N$ is indicated in the figure. In (a) the coupling to the leads is $U=0.75$ and
in (b) $U=0.50$.
By decreasing $U$ the shoulders in (a) turn to clear steps in (b). The bias is fixed to $V=2.0$, and $kT=0.0001$.}
\label{figure6}
\end{figure}

One observes that $G^R(t,s)=0$ for $s\geq t$ and, more interestingly, that $G^R(t,s)$
and $G^<(t,s)$ exhibit pronounced oscillations as time varies and reach a limit value 
as $s$ approaches $t$. In the case of the retarded Green function this limit is {\it constant} and equals
$-i$, which is simply the value of the unperturbed retarded Green function. This feature is easy to understand
by looking at the Dyson equation and noticing that when $s\to t$ the integration range of the inner integral
shrinks considerably so that at almost equal times the perturbed Green function resembles the unperturbed one.  
This argument is not restricted to the single site case we are discussing.
Also, since the spectrum of the discrete Laplacian is symmetric, the unperturbed retarded Green function
will always be real and therefore the real part of the full Green function will always be vanishingly small, 
as we shall check numerically in the many site case.  
In contrast, the limit of 
${\rm Im}G^<$ as $s\to t$ is {\it not} a constant w.r.t. $t$ but shows oscillations  
that  
disappear as $t$ increases. In the particular single-site case the limit value of $G^<$ is clearly the
occupation number of the site, whose oscillations were shown already in Fig.\,2(b).

 Figs.\,5 (a) and (b) show 3D maps of the imaginary part of the 
lesser Green function and emphasize the role of the coupling strength 
on the 
transient. At moderate coupling $U=0.75$ one observe small amplitude
oscillations except for $s\sim t$, in clear contrast to the case $U=1.20$ where pronounced 
oscillations exist even for large time differences. Inspecting the real part of the 
function $F_2$ given in  Fig.\,5 (c) we see that it does not depend on $t$ when
$s\sim t$ and that this gives the main contribution to the integral (\ref{partial}). 
It is now clear from  Figs.\,5 (a) and (c) that the corresponding current will be nearly stationary 
once the sample is charged (i.e. for $t>0.75$), because by increasing $t$ the 'off-diagonal' 
contributions are very small (some cancelations being possible as well). When $U$ increases the 
integral will collect instead 
nonnegligible contributions from the entire range $(0,t)$ and therefore the current will oscillate. 
These observations lead to the following statement: The steady state will be achieved 
at instant $t_s$ if for any $t>t_s$ there are no contributions for long-time differences, i.e.\ when both contact Green functions 
$G^R(t,s)$ and $G^<(t,s)$ vanish for $s>t_s$. It is easy to observe that this condition implies
the well known criteria for the steady state $G(t,s)=G(t-s,0)$. 

\subsection{Many-site case.}

\begin{figure}[tbhp!]
\includegraphics[width=0.45\textwidth]{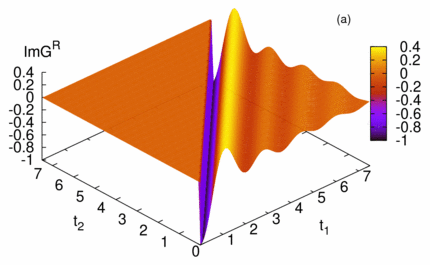}
\includegraphics[width=0.45\textwidth]{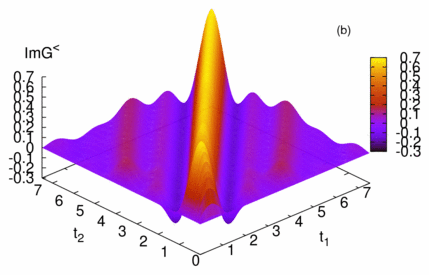}
\includegraphics[width=0.35\textwidth, angle=-90]{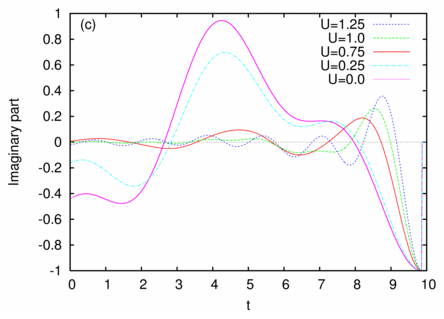}
\caption{(Color online) The imaginary part of the left contact retarded Green function (a) and of the 
lesser Green function
for the 4 site system (b). (c) The imaginary part of $G^R_{11}(t_1=10,t_2)$
 at different
couplings $U$. At small
coupling ${\rm Im}G^R$ resembles the retarded Green function of the isolated system (the curve corresponding
to $U=0.0$).}
\label{figure7}
\end{figure}

We consider now the more interesting case where the central region has more
than one level. 
Fig.\,6(a) emphasizes the qualitative differences between the
transients of 1D systems with $N=2,4,6,8$ sites. A general feature
is that as the size increases the transient develops a 'shoulder'
which is not met in the single site case.
For $N=2$ a second smaller slope of decrease
is noticed for $t\in [1.6:2.5]$. For $N=4$ the system experiences
few very different regimes before reaching the steady state. It first
decreases faster up to $t\sim 1.25$. For a short time a 
small hill develops around $t\sim 2$ then 
the decrease continues but clearly at a smaller rate. Finally, for
$t>3.25$ the current approaches the steady state very slowly. A similar
behavior is observed for $N=6$ and $N=8$, the main difference being that the
'shoulder' is longer and the intermediate slope is smaller. The patterns described above
suggest that there are some intermediate regimes, some of them being
characterized by a rather stable current. 
 Fig.\,6(b) emphasizes that at lower coupling $U=0.50$ the transient is even smoother and
for $N=4,6$ and 8 one notices the formation of clear steps.   

When the coupling strength $U$ is increased the transient shows oscillations but
they are fewer than in the single site (not shown).  
Since the two oscillatory integrals over energy in the current formula and both
self-energies do not depend on the number of sites in the system the
above size effects should be explained only by the behavior of the
contact Green functions. We show in Fig.\,7(a) and (b) the imaginary parts of the 
the contact Green functions $G^{R,<}_{11}$ of the 4 site system for a strong coupling $U=0.25$
(it turns out again that the real part of $G^{R}_{11}$ is vanishingly small).
Comparing with Fig.\,4 it is obvious that for the 4 site dot the Green functions 
have a more regular behavior and in particular the occupation number of the contact site 1
shows milder oscillations than the occupation number of the single site. 
Fig.\,7(c) gives the ${\rm Im}G^{R}_{11}(t_1=10,t_2)$ as a function of $t_2$
for different couplings and
reveals that at weak coupling to the leads the full retarded Green
function is close to the unperturbed one and the electron
dynamics inside the system must resemble the one of the isolated
sample.  
Indeed, the curve at $U=0.25$ follows
the oscillations of the free Green function which is also given in the figure (the curve corresponding
to $U=0.0$). We plot the imaginary
part since it turns out again that the real part is vanishingly small so
it does not contribute considerably to the retarded current. As $U$
increases the Green function changes and shows clear oscillations
imposed by the leads' self-energy. 

The analysis performed so far was focused on the behavior of the transient current 
as the intrinsic parameters of the system (i.e.\ its size and the height of the tunneling barriers
at the contacts are varied. Nevertheless, in a typical transport experiment these parameters are fixed and
one usually measures the current by varying the bias or a plunger gate voltage. 
From steady-state current measurements it is well known that the role of such a gate potential is 
to bring one or more levels of the quantum dot within the bias window (BW). 
We show in what follows that at fixed bias and given coupling strength to the leads one can tune 
the transient current with a gate potential. Moreover, by inspecting the transient behavior as
the gate potential is varied, it is possible to extract some information about the number of states 
within the bias window or above it. We will make the discussion for the 4 site dot. The gate potential 
is simulated by the diagonal term $V_g$ added to the on-site energy of the system. We fix the 
bias window to W=2.0 and for convenience we set $\mu_R=0.0$. Fig.\,8(a) give a families of 
transients for 
coupling strength $U=0.75$ and various values of $V_g$ specified in the figures.
Fig.\,8(b) shows the four levels of the isolated quantum dot as the gate potential scans 
the range $[-4:4]$.
The $V_g$ values chosen in Fig.\,8(a) corresponds to different location of the levels w.r.t. to BW.  
The bottom curve is irregular and settles down to a vanishing current because in this case there is no level 
within the BW. We note however that a nonvanishing transient current still develops shortly after the
coupling is established. At $V_g=-1.0$ the highest level is located in the BW and the transient is  smooth 
and already shows the additional shoulder
noticed previously. The same thing happens when two states lie in the BW (at $V_g=0.0$), the difference being that the
steady state current increases considerably. For $V_g>0.50$ it is clear from the structure of the spectrum that
one cannot have more than two states in the BW and that the levels pass gradually above it. 
We found interesting to look at the transient currents for those gate potentials that still allow      
two states in the transmission range while pushing one or two states above BW. One notices that 
for $V_g=1.0$ the steady state currents do not distinguish the different spectral structure involved in 
transport,
while the transient current is very sensitive to it. 

In the case of the six site QD the shoulder in Fig.\,4
 is 
more pronounced because at $V_g=0.0$ there are exactly three states inside the bias window.
We want to point out that since we have neglected the Coulomb interaction our simulations 
cannot capture the transport through excited states of the quantum dot. Tunneling processes involving such states 
would lead to a minipeak structure of the current maxima. We can consider however our results should describe
qualitatively the transport involving more levels because for small dots the bias required to cover the ground 
state of $N$ electrons is much higher that the excitation energies.  

\begin{figure}[tbhp!]
\includegraphics[width=0.45\textwidth]{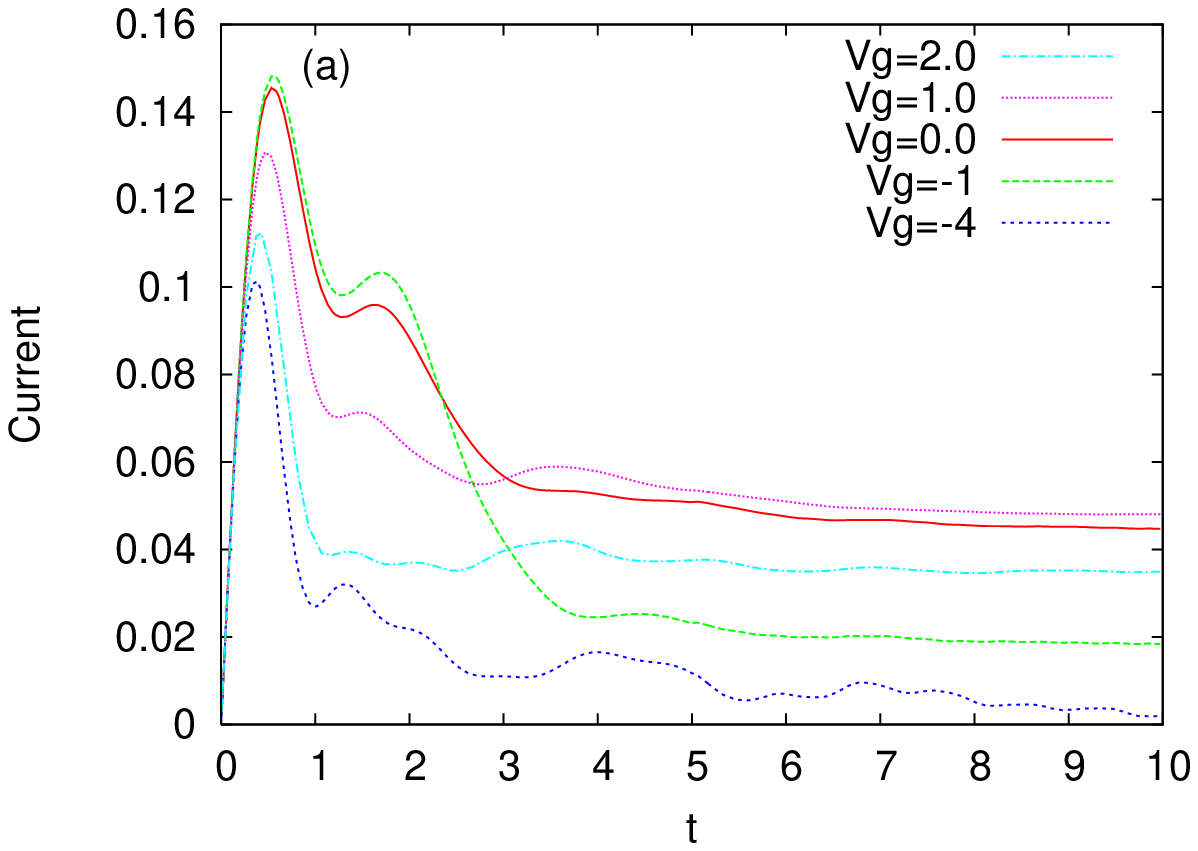}
\includegraphics[width=0.45\textwidth]{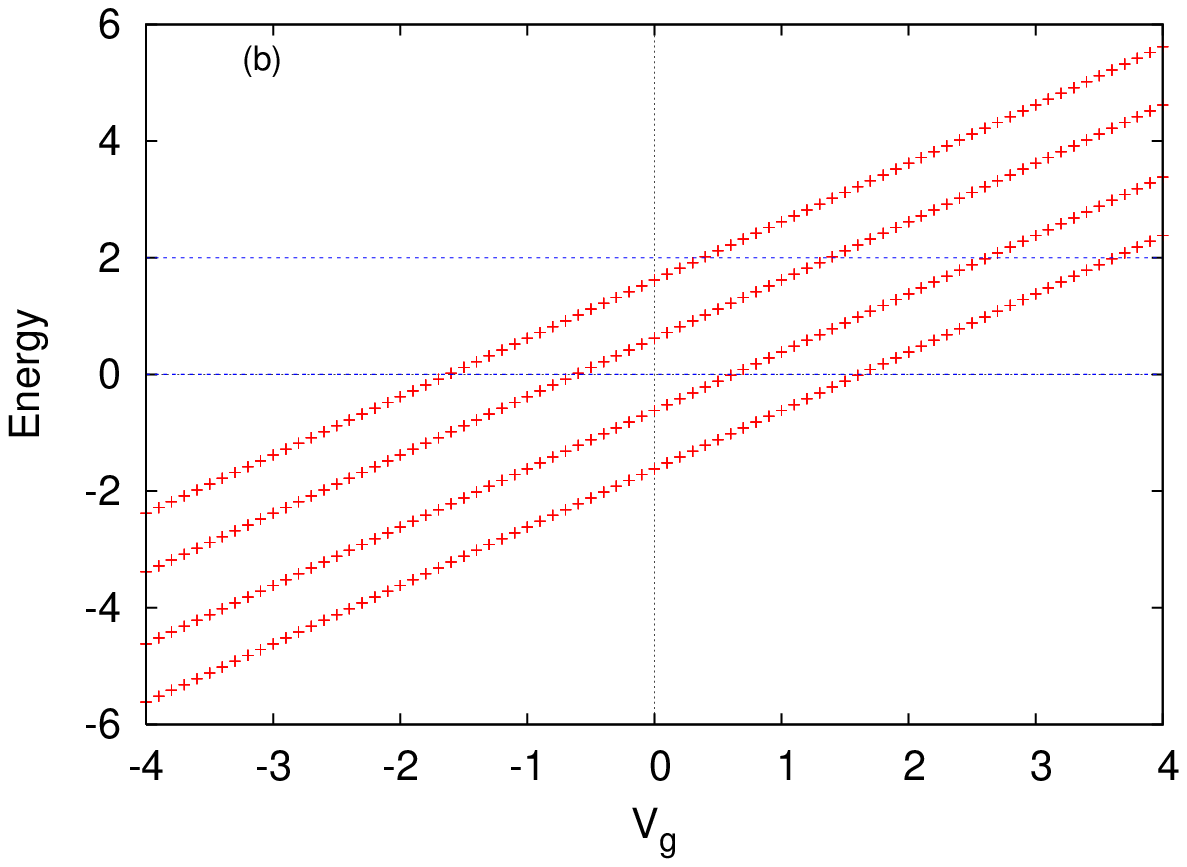}
\caption{(Color online) (a) The transient current for different values of the gate potential applied on the 4 site quantum dot for
 $U=0.75$. (b) The spectrum of the system as function of the gate potential.
It can be checked that the additional shoulders observed in (a) develop when there is at least
one state of the QD in the bias window and no state above it. Other parameters: $V=2$, and $kT=0.0001$.
Note that the bias is applied asymmetrically, that is $\mu_{\beta}=0.0$.}
\label{figure8}
\end{figure}

\begin{figure}[tbhp!]
\includegraphics[width=0.42\textwidth]{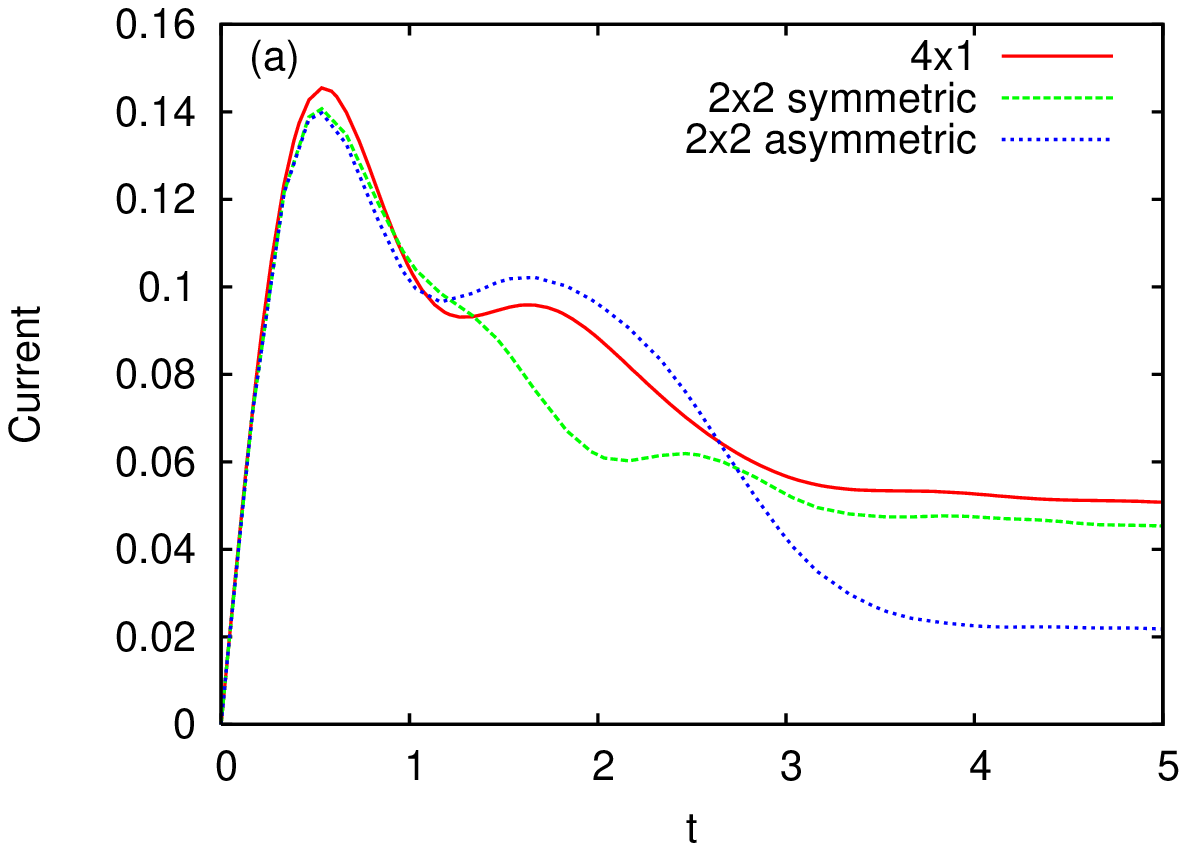}
\includegraphics[width=0.42\textwidth]{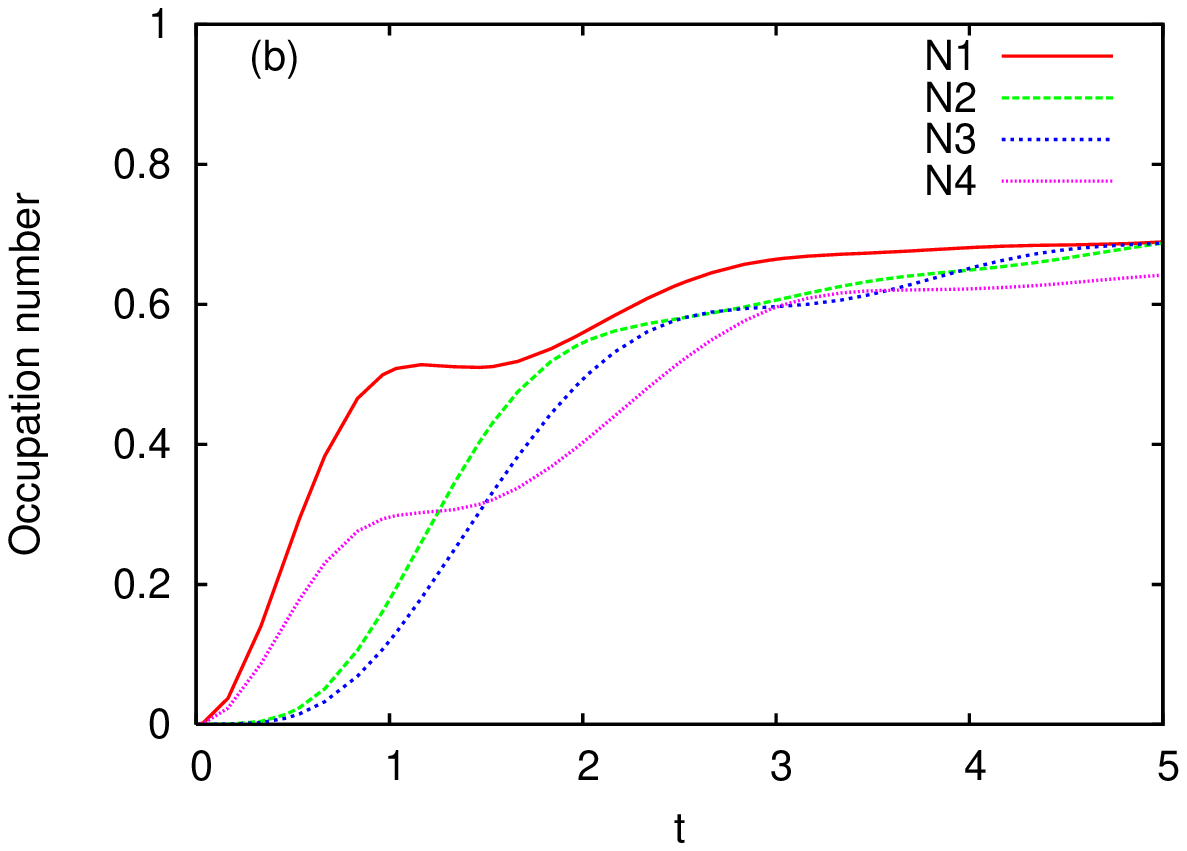}
\includegraphics[width=0.42\textwidth]{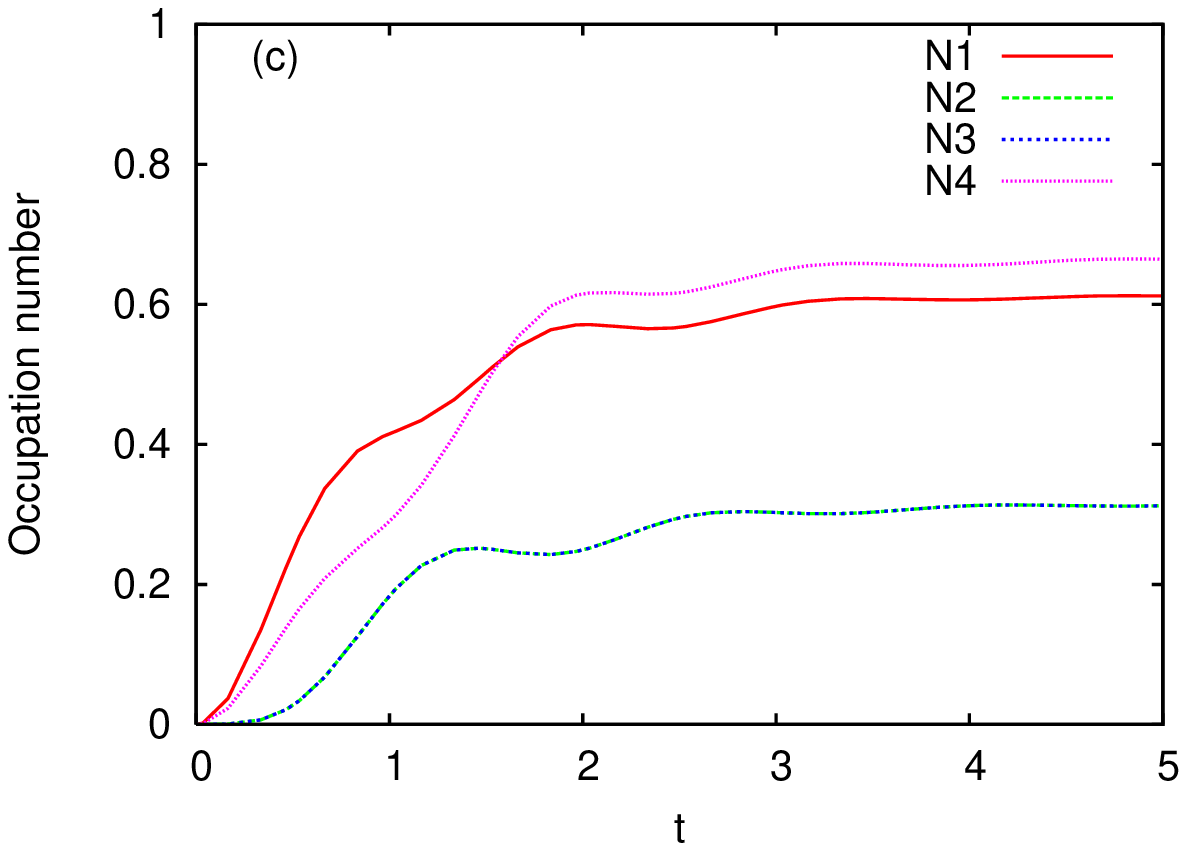}
\includegraphics[width=0.42\textwidth]{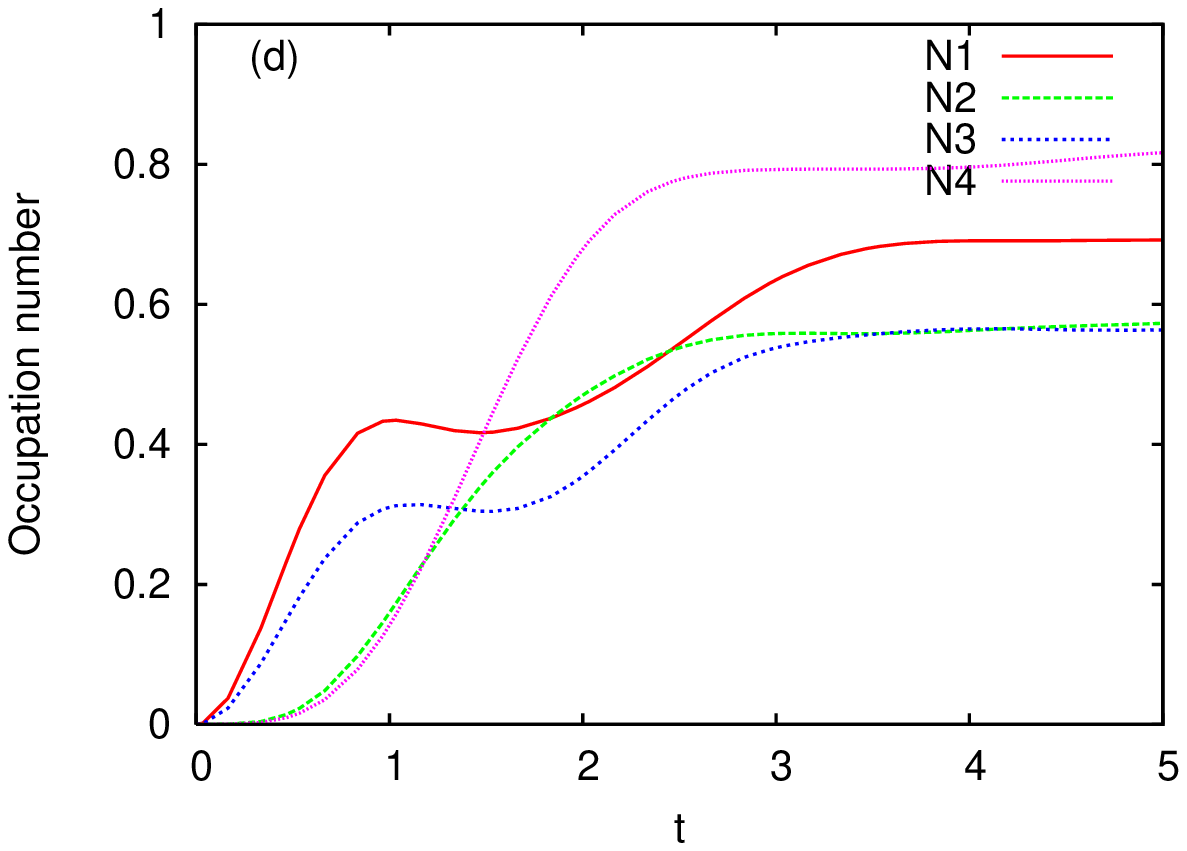}
\caption{(Color online) (a) The transient in the left lead $J_{\alpha}(t)$ for the $4\times 1$ system and for the $2\times 2$ system in
the two configurations of the leads as mentioned in the text. (b) - 1D system, (c) - symmetric configuration, 
(d) - asymmetric The occupation numbers $N_i(t)$ of
the $i$-th site for the three cases considered in Fig.\,8(a).
Other parameters: $U=0.75$, $V=2.0$, and $kT=0.0001$.}
\label{figure9}
\end{figure}

\begin{figure}[tbhp!]
\includegraphics[width=0.45\textwidth]{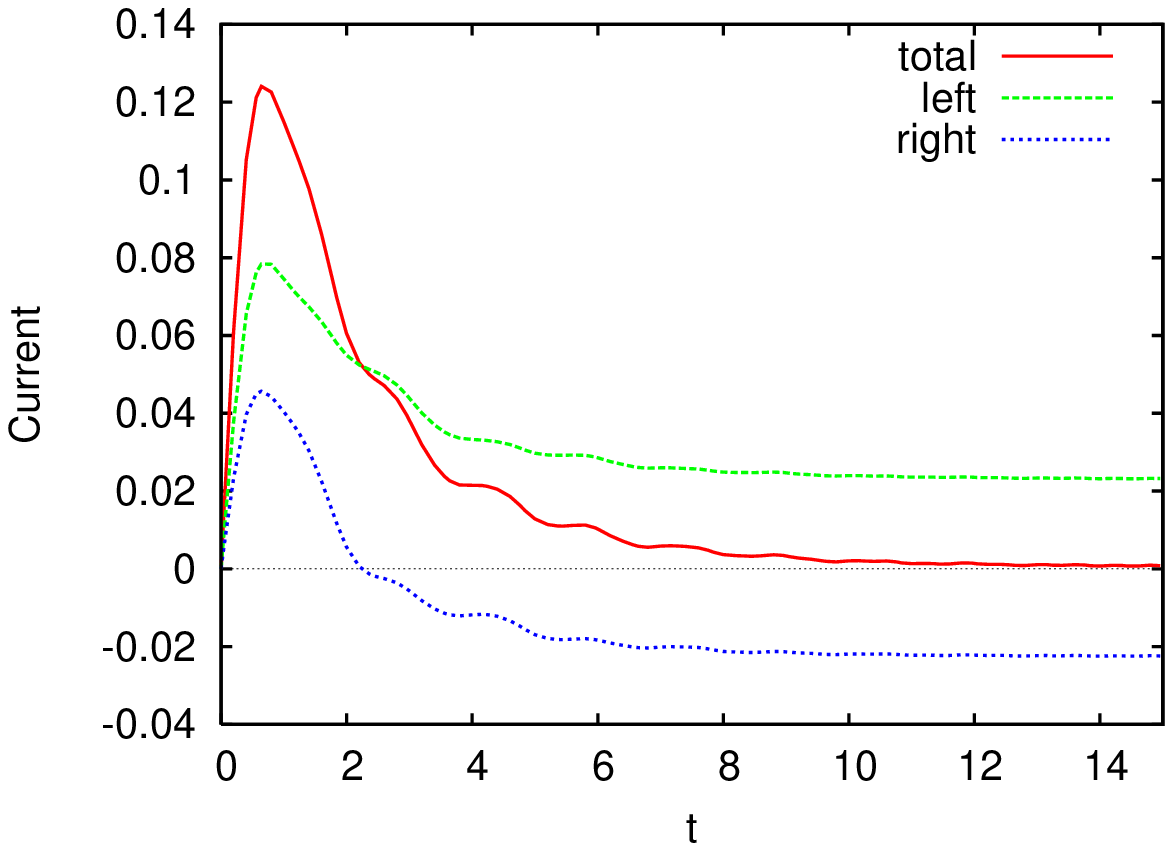}
\caption{(Color online)
 The total transient current and the components in the left lead and right lead.
Other parameters: $U=0.5$, $V=2$, $kT=0.0001$.}
\label{figure10}
\end{figure}

Now we investigate two more features of the transient regime: The time-dependent charge  
filling of the central region and possible effects due to different shapes of the system or 
of the various ways in which one can couple the leads. Besides the 4 site 1D system discussed
so far we consider also a $2\times 2$ quantum dot. Both systems are submitted to the same bias and have
equal coupling to the leads. However, in the case of the 2D quantum dot one can use different
contacts for plugging the leads. We discuss two situations: i) A symmetric configuration in which 
the leads are attached to the opposite
corners of the system, namely at the sites 1 and 4 and ii) An asymmetric coupling when we use the 1st
and 3th sites as contacts. Fig.\,9(a) reveals the changes induced in the transient curves  
in each case. The remaining subfigures show the occupation number   
$N_i(t)$ of each site $i=1,4$ for the 1D system and the two configurations considered. 
Inspecting Fig.\,9(a) and (b) we
see: i) The contact sites 1 and 4 are the first to be populated due
to their proximity to leads; ii) Since $\mu_{\alpha}>\mu_{\beta}$
the right contact site (the 4-th) gains less charge at a lower rate
than the left contact (the first). Both $N_1$ and $N_4$ show a
step-like behavior for a short period (around $t=1$). 
This coincides with the increased occupation number on the middle sites.
We note also that the step in the occupation of the contact site $N_1$ ends
when it is equaled by $N_2$.  
All the sites are then continuously filled up to the steady-state
value. The occupation number on the right contact is smaller than
the other ones which attain roughly the same value $0.65$. 
iii) The step-like behavior of the transient currents in
the range $[1:1.75]$ corresponds to the almost constant population
of the contact sites in the same interval.
The symmetric configurations is still characterized by a smooth transient
but we notice that the step appears now later that in the 1D case. Fig.\,9(c) 
confirms again that this stable regime is assigned to a constant flow in the 
contact site. Also it reflects that the charge is equally distributed in the 
sites 2 and 3 which are located symmetrically w. r.t. the leads.
In the asymmetric geometry the transient is rather similar to the one of the 1D
system up to $t=2.8$ but then drops to a lower steady state value. The occupation numbers 
show that the fourth site carries more charge than the contact sites in the steady state. 
This means in our opinion that part of this charge simply accumulates and is not participating in transport. More interestingly, we note that in
contrast to the symmetric geometry $N_2$ and $N_3$ are different in the transient
regime but reach the same value in the steady state. 
We mention that time-dependent simulations were performed recently in the case of an 
Aharonov-Bohm ring starting from the Schr\"{o}dinger equationi.\cite{Peeters}

We discuss now briefly the total current $J_{\alpha}+J_{\beta}$ which is given in Fig.\,10
along with its two components
(the $+$ sign is due to the fact that the current $J_{\beta}$ represents the
current from the lead $\beta$ to the system and therefore has opposite sign).
As already mentioned, only in the steady state the current conservation reads as
$J_{\alpha}=-J_{\beta}$. Consequently, the net current $J_{\alpha}+J_{\beta}$ vanishes.
However, in the transient regime the two currents, although having similar shape
differ significantly. 

\begin{figure}[tbhp!]
\includegraphics[width=0.45\textwidth]{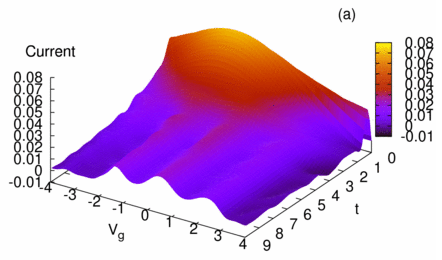}
\includegraphics[width=0.45\textwidth]{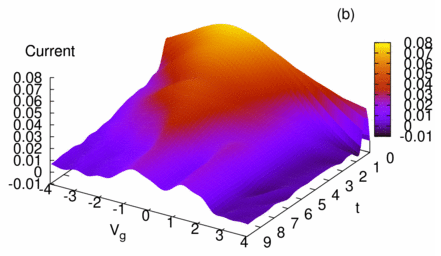}
\includegraphics[width=0.45\textwidth]{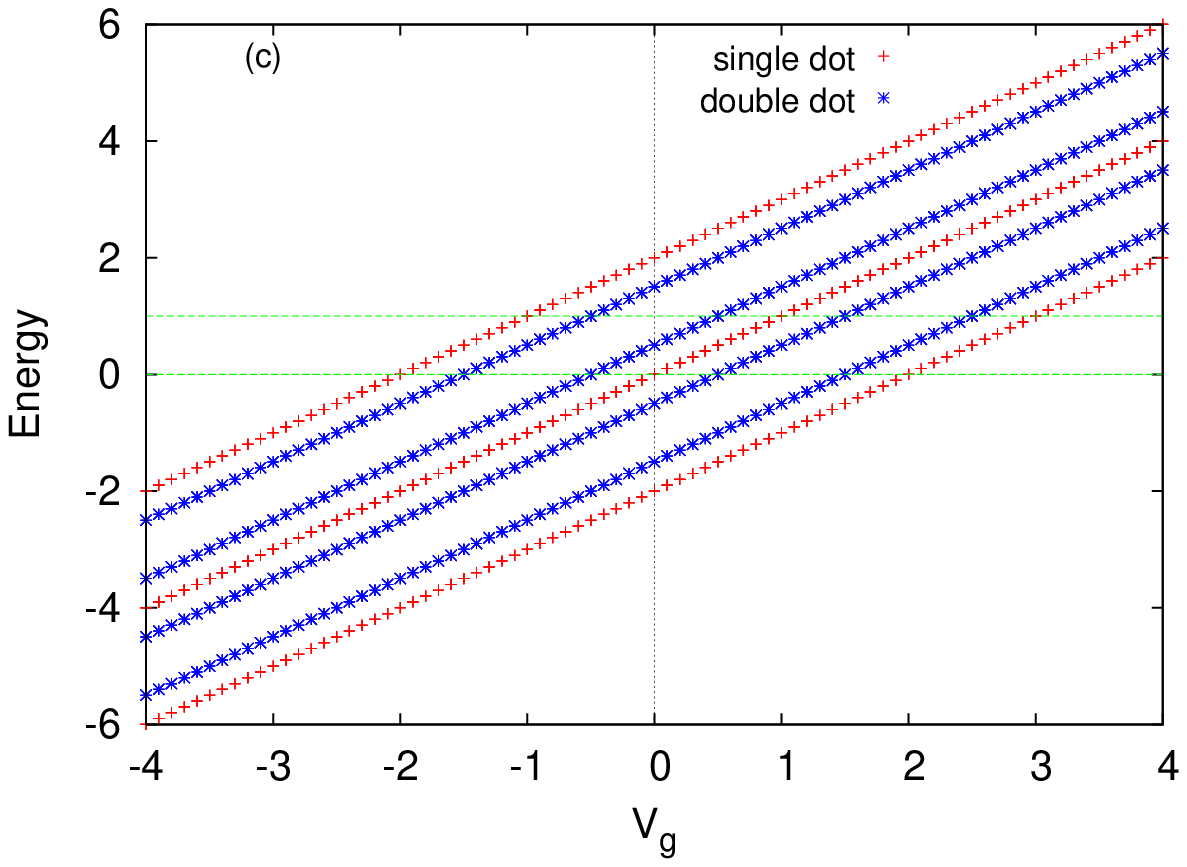}
\vskip 0.5cm
\caption{(Color online) (a) The 3D map of the transient current for a $2\times 2$ site QD
as a function of time and gate potential $V_g$.
The maxima are related to the spectrum of the isolated system given in (c).
(b) The 3D map of the transient current for $2\times 2$ the double dot $t_{24}=t_{13}=0.5$
Other parameters: $V=1.0$, $U=0.50$, $kT=0.0001$. (c) The spectra of the two systems as a function of the
gate potential. The lines mark the bias window. }
\label{figure11}
\end{figure}

To substantiate further the previous analysis we present in  Fig.\,11(a) the contour plot of the current
as function of time and gate potential for the $2\times 2$ site quantum dot coupled to the leads in the
symmetrical configuration. Fig.\,11(c) gives the spectrum of the system as the gate potential varies. 
The middle eigenvalue is doubly degenerated.
When the levels are either below or above the
bias window ($W=1.0$ starting from $\mu_R=0.0$) the transient oscillates for quite a long time 
before passing to the steady state.
We also observe that in these two extreme limits the transient oscillations are qualitatively different.
For $V_g\sim -4$ the current shows decreasing oscillations towards the steady state. In contrast, for
$V_g\sim 4$ one remarks faster oscillations and more importantly, negative values of the transient.
Since in this regime all the levels are above the bias window and there is no way to pass
electrons to the right
lead it is clear that the negative current in the left lead is just the reflected one. We underline
that this effect is due to the fact that we have considered a finite spectral width of the leads and
that similar features were reported for the single site case \cite{Maciejko}. As expected, as the
system approaches the stationary regime the current shows three maxima associated to the passage 
of the localized levels through the bias window. Actually the levels turn to resonances when coupling the
leads to the system but since one has to deal with a time-dependent Hamiltonian it is difficult
to characterize the location and width of these resonances in the transient regime. 
This is why in the 3D plot one cannot distinguish between different resonances at times $t<3.0$.  
Fig.\,11(b) presents the transient current for the same $2\times 2$ system except that the hopping parameters
$t_{13}$ and $t_{24}$ are reduced to 0.5. In this case one can view the system as a double dot, each dot composed 
of two sites. As the spectrum from Fig.\,11(b) shows, the degeneracy is lifted and the level spacing diminishes. 
As a consequence in the long time regime one gets two broader peaks, since the four levels are now grouped into pairs. 

All the features presented above emphasize that the transient regime of the many-level structures 
is quite different from the single-level system.

\section{Conclusions}

We have performed transient current calculations for a many-level
finite system coupled suddenly to semiinfinite biased leads. Our
method is based on the nonequilibrium Green-Keldysh machinery. We
find numerically an exact solution of the integral Dyson equation
which is solved as an algebraic equation. By analyzing the behavior
of the retarded and lesser Green functions we explain qualitatively
the shape of the transient current and the passage to the steady
state. 
The amplitude of the coupling to the leads controls essentially the
convergence to a steady state.
We have identified non-trivial effects of the many-level
structure of the system and presented an intuitive picture of the
charge filling by studying the occupation number inside the system.
By increasing the
system size the shape of the transient current and the evolution towards the 
steady state differs significantly
from the single-site oscillatory behavior and depends crucially on the number of 
electronic states available in the bias window. We predict that a step-like structure
could be observed in transient current measurements by applying a gate 
potential on the system that 
tunes the higher
levels within the bias window. Different transients are expected to appear as well when 
different coupling geometries of the leads are used.    

The present method can be used for studying the response of mesoscopic systems to
more complicated time-dependent couplings to the leads: pulses having different lengths and
decaying rates, and non-periodic signals.

\acknowledgments{This work was supported in part by the research programme of the Icelandic Reaserch
Council for Nanoscience.
 V.M was also supported by CEEX Grant D11-45/2005. We acknowledge useful discussions with C. S. Tang.}

\end{document}